\documentclass[prd]{revtex4}
\usepackage{graphicx}
\usepackage{dcolumn}
\usepackage{bm}
\usepackage{epsfig}

\begin{document}

\title{A Numerical Approach to Space-Time Finite Elements for 
  the Wave Equation}

\author{Matthew Anderson$^{1,2}$, Jung-Han Kimn$^{2,3}$}

\affiliation{$1$ Department of Physics and Astronomy, Louisiana State
University, Baton Rouge, LA 70803-4001\\
$2$ Center for Computation and Technology, Louisiana State
University, Baton Rouge, LA 70803\\
$3$ Department of Mathematics, Louisiana State
University, Baton Rouge, LA 70803-4918}

\begin{abstract}
We study a space-time finite element approach
for the nonhomogeneous wave equation
using a continuous time Galerkin method.  We present
fully implicit examples in $1+1$, $2+1$, and $3+1$ dimensions using 
linear quadrilateral, hexahedral, and tesseractic elements.  
Krylov solvers with additive Schwarz preconditioning are used for solving 
the linear system.  We introduce a time decomposition 
strategy in preconditioning which significantly improves performance
when compared with unpreconditioned cases.
\end{abstract}
\maketitle

\section{Introduction}
Space-time finite elements provide some natural advantages for numerical relativity.  With space-time elements, time-varying computational 
domains are straightforward, higher-order approaches
are easily formulated, and both time and spatial domains can be
discretized using a single unstructured mesh.  However, while continuous
Galerkin approaches employing space-time finite elements have found 
use in many engineering 
applications \cite{Csik}, \cite{Kim}, \cite{Idesman}, \cite{Guddati}, \cite{Kit},  
  they have not been used in numerical relativity.
Recent numerical relativity evolutions using finite elements employ
discretization of the space domain and marching in time rather than
simultaneous discretization of both space and time domains
\cite{Sopuerta}, \cite{Cherubini}, \cite{Cherubini2}.

We investigate a space-time finite element method similar to \cite{French1996}
using continuous approximation functions in both space and 
time to explore its use for 
numerical relativity simulations.  
The main purpose of this paper is to present our numerical results. 
We present a time-parallel preconditioning strategy for use with continuous space-time
elements and Krylov solvers, and explore 
numerical results in $1+1$ dimensions and higher.

Many space-time approaches to the wave equation exist (see \cite{Falk}, \cite{Hughes}, \cite{French91}, \cite{Hulbert}).   
Our approach is different in that we do not use time slab finite elements, which are continuous
in a limited domain of time (the time slab) but discontinuous between neighboring time slabs.
Instead, we discretize space and time together for the entire domain 
using a finite element space which does not discriminate between space and time basis functions
and consider iterative solution
methods with a time decomposition preconditioner. This approach has advantages for 
more general finite element spaces and parallelization.  In 
this paper, however, we restrict ourselves to structured space-time finite elements and 
present the results obtained on a single processor 
in order to better compare results and performance with other approaches to solving the wave equation.

We consider the following nonhomogeneous wave equation problem with the initial and boundary value problem
to find $u({\bf x},t)$ such that
\begin{eqnarray}
\label{eqn:wave}
\frac{\partial^2 u}{\partial t^2} - \triangle u =  f \hspace{0.15cm} &{\rm in}& \hspace{0.15cm} \Omega \times [0,T], \\ \nonumber
u = u^0 \hspace{0.15cm} &{\rm on}& \hspace{0.15cm} \Omega \times \{t=0\}, \\ \nonumber
u_t = v^0 \hspace{0.15cm} &{\rm on}& \hspace{0.15cm} \Omega \times \{t=0\}, \\ \nonumber
u_n = 0 \hspace{0.15cm} &{\rm on}& \hspace{0.15cm} \partial \Omega \times [0,T],
\end{eqnarray}
where $\Omega$ is a bounded domain in $R^d, \,\, d =1,2,3$ and $u_n$
is the outward pointing normal derivative. 
As in \cite{French1996}, Eq. (\ref{eqn:wave})
is re-written to be first order in time by introducing an auxiliary variable, $v=u_t$:
\begin{eqnarray}
\frac{\partial v}{\partial t} - \triangle u = f \hspace{0.15cm} &{\rm in}& \hspace{0.15cm} \Omega \times [0,T],  \\ \nonumber 
-\frac{\partial u}{\partial t} + v = 0 \hspace{0.15cm} &{\rm in}& \hspace{0.15cm} \Omega \times [0,T],  \\ \nonumber
u = u^0 \hspace{0.15cm} &{\rm on}& \hspace{0.15cm} \Omega \times \{t=0\}, \\ \nonumber
v = v^0 \hspace{0.15cm} &{\rm on}& \hspace{0.15cm} \Omega \times \{t=0\}, \\ \nonumber
u_n = 0 \hspace{0.15cm} &{\rm on}& \hspace{0.15cm} \partial \Omega \times [0,T],\\ \nonumber
v_n = 0 \hspace{0.15cm} &{\rm on}& \hspace{0.15cm} \partial \Omega \times [0,T].
\end{eqnarray}
We use a nonhomogeneous Dirichlet boundary condition on the 
initial boundary $\Omega \times \{t=0\}$,  and  
a homogeneous Neumann boundary condition for $\partial \Omega \times (0,T]$. 
No boundary condition is set at $\Omega \times \{t=T\}$ to avoid
overspecifying the problem.  Consequently, the 
evolution equations themselves become
an effective boundary condition by determining the values
for the solution at $\Omega \times \{t=T\}$.

The space $L^2(\Omega)$ is defined as the closure of $C^{\infty}(\Omega)$ in the norm,
\[ ||u||_{L^2(\Omega)} = \left( \int_{\Omega} |u|^2 dx \right)^{1/2} < \infty. \]
The $H^1$-seminorm and norm of $u \in H^1(\Omega)$ are, respectively,
\[ |u|^2_{H^1(\Omega)} =  \int_{\Omega} |\nabla u|^2 \,dx; \quad ||u||^2_{H^1(\Omega)} =  |u|^2_{H^1(\Omega)} + ||u||^2_{L^2(\Omega)}. \]
We define the Hilbert space $L^2([0, T],H^1(\Omega))$ by
\[ ||w||_{ L^2([0, T],H^1(\Omega))}= \Big(\,\, \int^T_O ||w(\cdot, t) ||^2_{H^1(\Omega)} \, dt \,\,\Big)^{1/2}. \]
For the space-time finite element space of $n=1,2,3$ spatial dimensions, 
  we consider the standard finite element space of $n+1$ dimensions. 
  Therefore our finite element space  $V$ is the space of piecewise polynomial 
  functions $\phi: \Omega \times (0, T] \rightarrow R$. 

The weak form is to find approximate solutions $\tilde{u},\tilde{v} \in L^2([0, T],H^1(\Omega))$ such that
\begin{eqnarray}
M\left(\tilde{u},\tilde{v},\phi\right) & = & 0 \hspace{0.15cm}  \label{eqn:A}\\
N\left(\tilde{u},\tilde{v},\phi\right) & = & 0 \hspace{0.15cm} \hspace{0.35cm} \forall \phi \in L^2([0, T],H^1(\Omega)), \label{eqn:B}
\end{eqnarray}
where
\begin{eqnarray}
M\left(\tilde{u},\tilde{v},\phi \right) & = & \int_{\Omega \times [0,T]} \left( \frac{\partial \tilde{v}}{\partial t} \phi + {\bf \nabla}\tilde{u} \cdot {\bf \nabla} \phi - f \phi \right) ds \label{eqn:C}, \\
N\left(\tilde{u},\tilde{v},\phi \right) & = & \int_{\Omega \times [0,T]} \left( -\frac{\partial \tilde{u}}{\partial t} \phi + \tilde{v} \phi \right) ds.\label{eqn:D}
\end{eqnarray}

Motivated by the success of domain decomposition methods for general sparse matrices \cite{Toselli:2004:DDM}, \cite{Sarkis}, \cite{Cai}, 
we also examine additive Schwarz methods \cite{6wid}, \cite{2dry}, \cite{6dry}, \cite{1cs}, \cite{Xu} with a time decomposition preconditioning strategy. 
While additive Schwarz preconditioning has been applied to hyperbolic problems 
before \cite{Wu}, \cite{Cai2}, applying additive Schwarz preconditioning to space-time
elements using a time decomposition strategy is unique to this work.

\section{Numerical Results}
In this section we present solutions to the nonhomogeneous wave equation using space-time elements in various
dimensions.  We use uniform structured meshes to better compare results with other approaches to solving the wave equation.
Solutions presented are produced by a single linear solve of the system in Eqs.~(\ref{eqn:A})--(\ref{eqn:B}). 
All codes presented use PETSc
\cite{petsc-web-page}, \cite{petsc-user-ref}, \cite{petsc-efficient}; the linear solve residuals 
given (labeled ``Final Residuals") are the absolute residual norms,
\begin{eqnarray}
r = || A x - b ||_{L_2}
\end{eqnarray}
for the linear system $ A x = b$ where $A$ is the system matrix, $x$ is
the solution, and $b$ is the system right hand side vector for both
$\tilde{u}$ and $\tilde{v}$.
We use the $L_\infty$ norm for reporting differences between the analytic and approximate solution:
\begin{eqnarray}
|| e ||_{L_\infty} = \max | e^i |, 
\end{eqnarray}
for vector $e$.  For Krylov solve examples, the initial guess given for the solution is always zero.
\subsection{$1+1$ Dimensions}
For $1+1$ dimensions, we consider the nonhomogeneous wave equation with solution
\begin{eqnarray}
U_e \left(x,t\right) = \exp\left[-\left(x-\cos t\right)^2\right] \label{eqn:1+1}
\end{eqnarray}
on a domain of $x = \left[-5,5\right]$ and $t=\left[0,10\right]$.
We choose the appropriate source term, $f$, in Eq.~\ref{eqn:wave}
\begin{eqnarray}
 f & = & -2\,\left(\cos t\right) {e^{- \left( x-\cos t \right) ^{2}}}
       \left( 2\, \cos^3 t -4\,x
         \cos^2 t + 2\,x^2\cos t 
         -2\,\cos \,t +x \right),
\end{eqnarray}
and initial conditions to produce this test problem solution.
Solving this system via LU decomposition
with linear rectangular elements we observe the expected second order convergence,
  shown in Table \ref{table:1p1_lu}.
\begin{table}
\begin{tabular}{c|c|c|c}
$n_x$ & $n_t$ & $\| \left(\tilde{u}-U_e\right) \|_{L_\infty}$ & rate \\
  \hline
60 & 60 &  $2.21 \times 10^{-2}$ &  --   \\
120 & 120 & $5.11 \times 10^{-3}$ & $2.08$ \\
240 & 240 & $1.26 \times 10^{-3}$ & $2.01$ 
\end{tabular}
\caption{Convergence using LU decomposition of a space-time element simulation with solution
  given by Eq. (\ref{eqn:1+1}).  There are $(n_x-1)(n_t-1)$ total elements in the mesh.
    The number of nodes in the $x$ and $t$ directions are $n_x$ and $n_t$, respectively.
Using linear elements, we expect second order convergence in 
the $L_\infty$ norm.  
The convergence rates reported 
are given by ${\rm rate} = \ln \left(E_2/E_1\right)/\ln \left(h_2/h_1\right)$ 
where $h_1,h_2,E_1,E_2$ are the successive quadrilateral lengths and
$L_\infty$ norms, respectively.}
\label{table:1p1_lu}
\end{table}
Since scaling with problem size using LU decomposition for a banded
matrix is $O \left( N b^2 \right)$ -- where $N$ is the size of the matrix and 
$b$ is the bandwidth
 -- LU is entirely inadequate for large problems with space-time elements.  
Krylov solvers \cite{Saad}, like GMRES \cite{Saad1986}, are much more suitable for such problems.

We tested a variety of solvers and preconditioners available 
in PETSc \cite{petsc-web-page}, \cite{petsc-user-ref}, \cite{petsc-efficient} for
the problem using a $60^2$ structured mesh.  The results are summarized in Table \ref{table:1d}.
\begin{table}
\begin{tabular}{c|c|c|c|c}
Solver Type & Preconditioner & iterations & Final Residual & $\| \left(\tilde{u}-U_e\right) \|_{L_\infty}$ \\
  \hline
GMRES & none & 5127   & $10^{-5}$ & $2.20 \times 10^{-2}$ \\
GMRES & none & 2000   & $10^{-2}$ & $3.68 \times 10^{-1}$ \\
LSQR  & --   & 5000   & $10^{-3}$ & $2.48 \times 10^{-1}$ \\
GMRES & Jacobi & 6000 & $10^{-2}$ & $8.68$ \\
GMRES & Block-Jacobi  & 1 & $10^{15}$ & -- \\
\end{tabular}
\caption{Various linear solver tests to solve the $1+1$ problem on a $60^2$ structured
mesh.  GMRES performed the best, but required a very large number of iterations.  LSQR
is similar to a direct method and cannot be preconditioned in PETSc.  Jacobi and Block-Jacobi
preconditioning made GMRES convergence even worse than unpreconditioned.}
\label{table:1d}
\end{table}
While GMRES converges without preconditioning, it requires a high number of iterations to obtain a physically 
meaningful result.  
Preconditioning with Jacobi or Block-Jacobi does not improve the convergence rate.  However, neither Jacobi
nor Block-Jacobi preconditioning offer much flexibility with respect to the geometry of the problem.  Additive Schwarz
offers more flexibility in preconditioning this hyperbolic problem.

\begin{figure}
\begin{center}
\epsfig{file=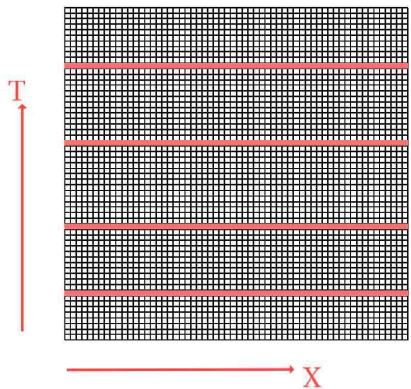,height=5.5cm}
\caption{The $60^2$ mesh used for the $1+1$ dimension simulations.  Here the entire mesh is
  divided into five subdomains in time for use in preconditioning.  The linear systems
    resulting from each subdomain are solved and the result used for preconditioning the
    global linear system.  
   The subspace interface condition is the same as for the original boundary conditions: 
   Dirichlet for $t = \left( t_{n-1} - {\rm overlap} \right)$ and evolution equation 
   determined for $t = \left( t_n + {\rm overlap} \right)$ where $t_n$ is the
   $n$-th time decomposition.
    This preconditioner is also time-parallel: each time subdomain
    could be solved simultaneously on a different processor.  Spatial domain decomposition
could also be applied, but we only examine time decomposition here.}
\label{fig:time_dd}
\end{center}
\end{figure}

We follow a time decomposition strategy for additive Schwarz, as illustrated in Figure \ref{fig:time_dd}.
The domain of the problem is split into separate subdomains of time slabs.  Each subdomain overlaps the face
cells of its neighbors.
The much smaller linear system of each subdomain is subsequently solved, either by GMRES or by LU decomposition, 
and the result used for preconditioning the global system.  
We respect the original boundary conditions for the subspace interface condition: 
Dirichlet for $t = \left( t_{n-1} - {\rm overlap} \right)$ and evolution 
equation determined 
for $t = \left( t_n + {\rm overlap} \right)$ where $t_n$ is
the $n$-th time decomposition.

\begin{figure}
\begin{tabular}{cc}
After 1 iteration & After 10 iterations \\
\epsfig{file=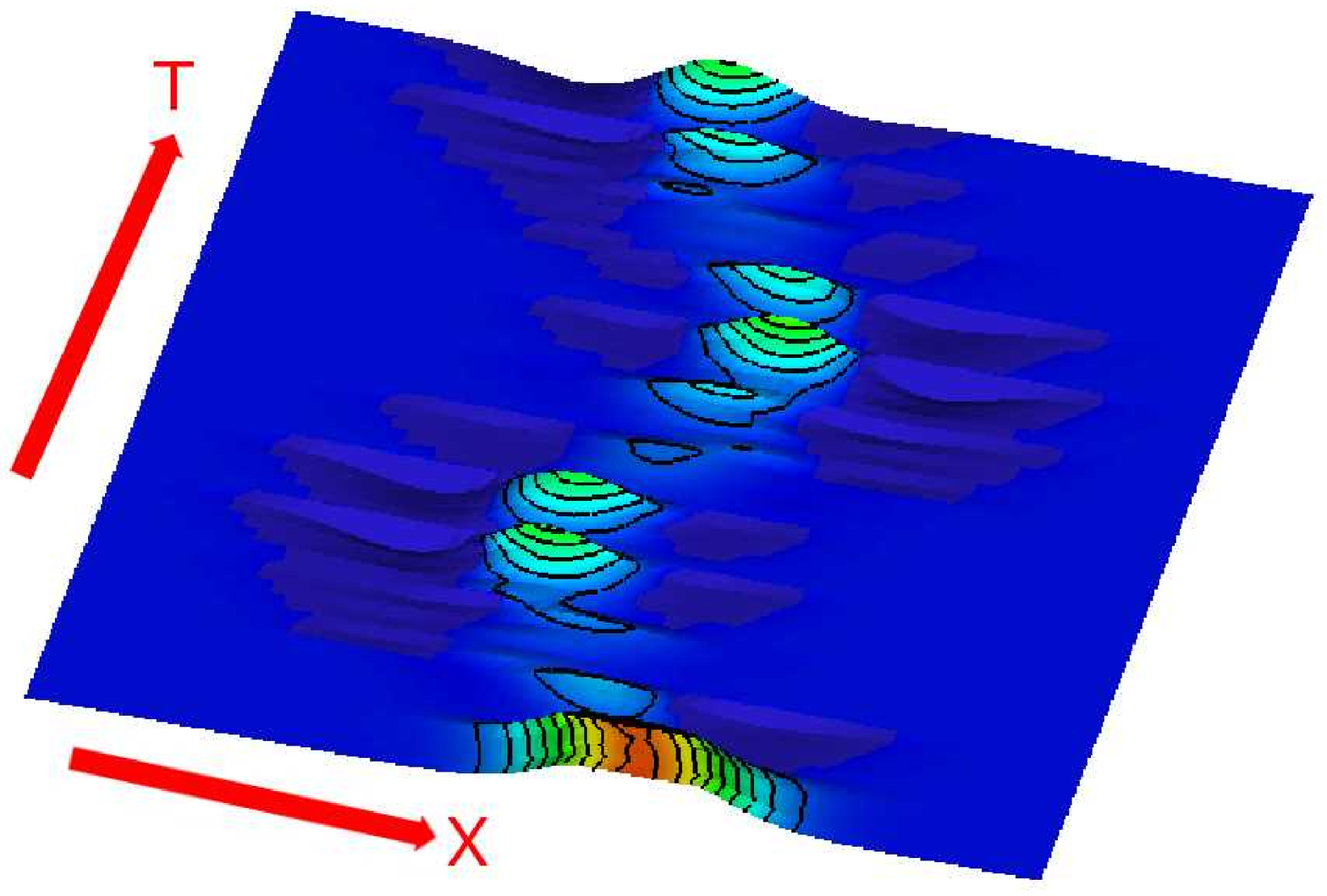,height=5.5cm} & \epsfig{file=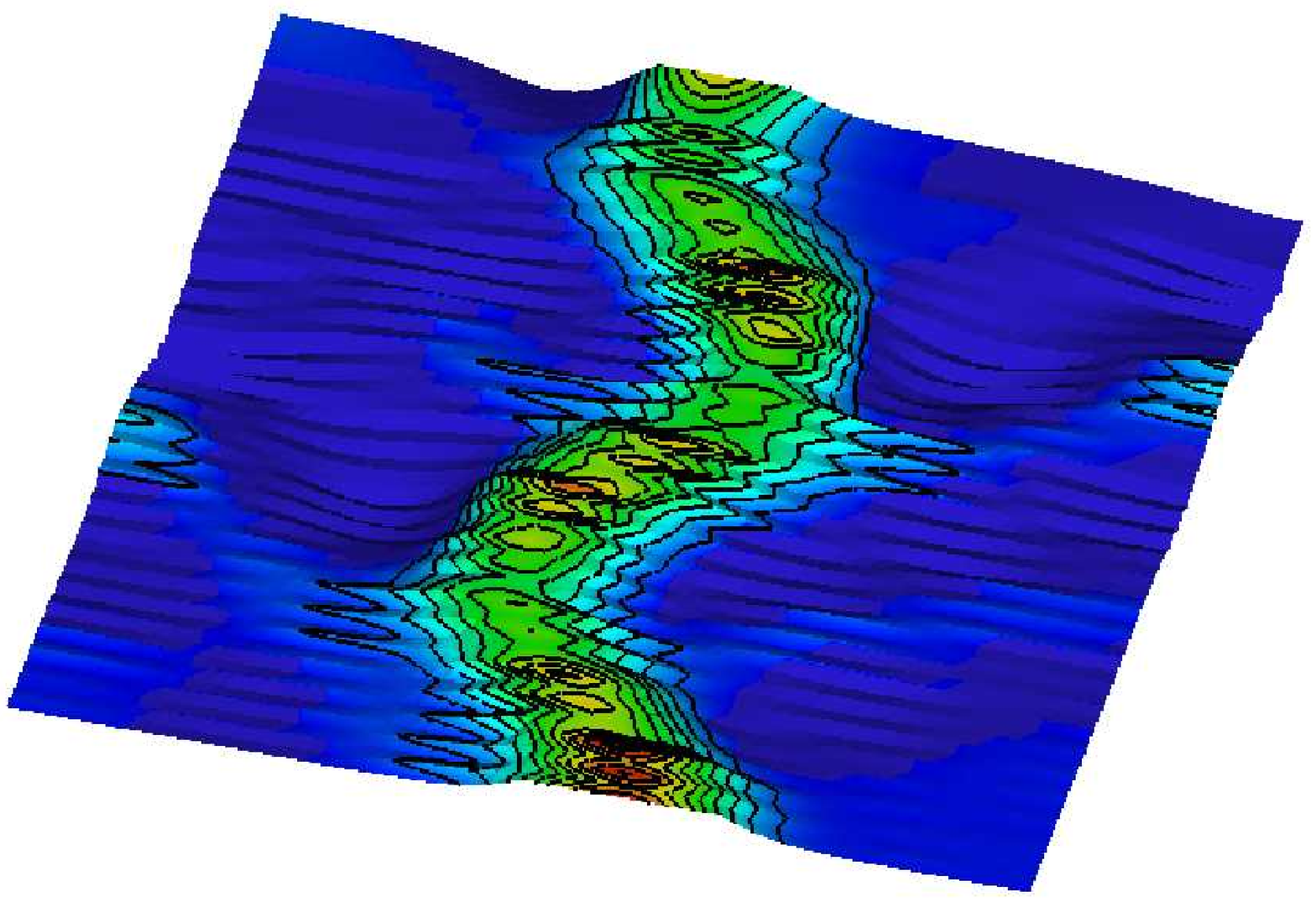,height=5.5cm} \\
 & \\
After 100 iterations & After 500 iterations \\
\epsfig{file=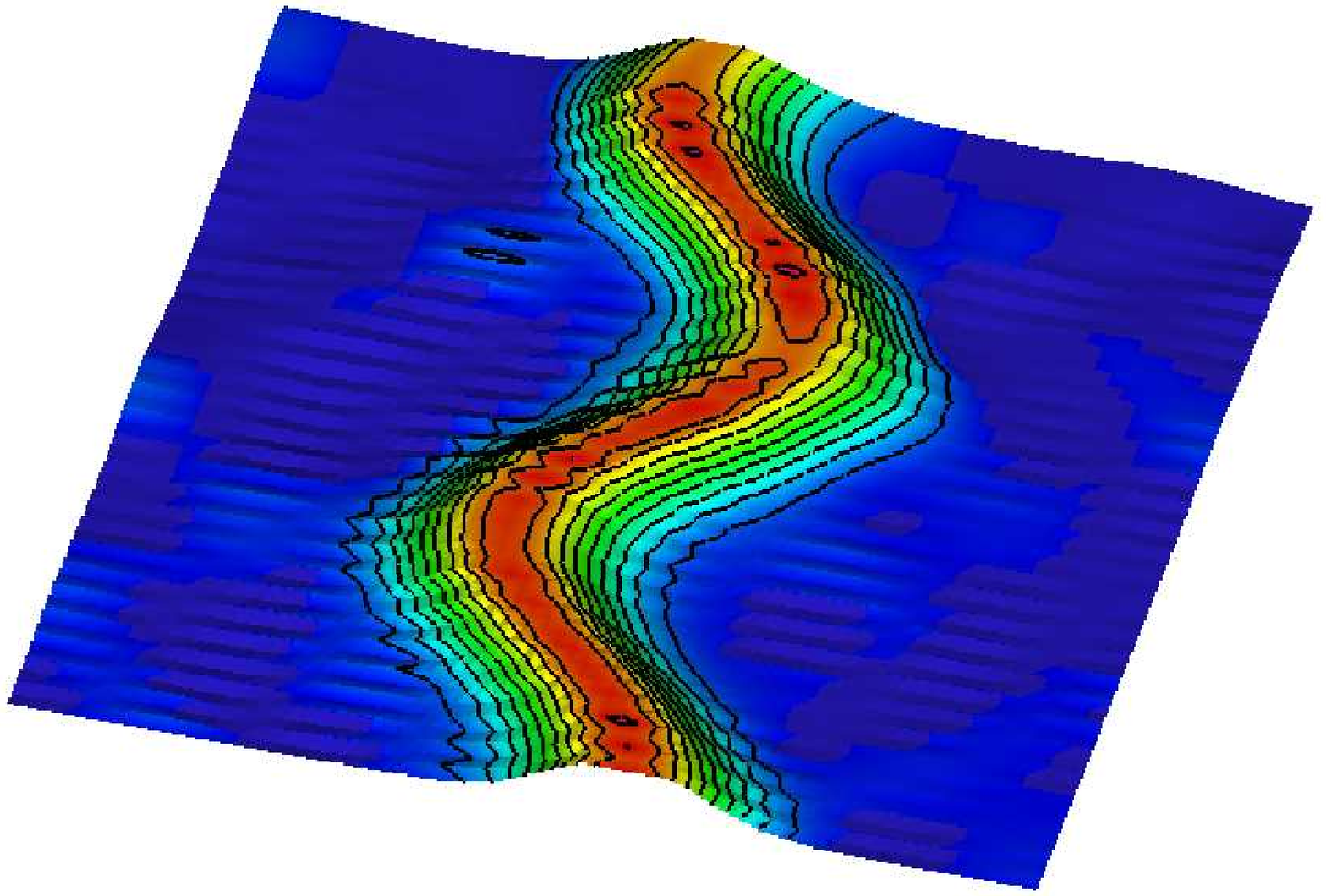,height=5.5cm} & \epsfig{file=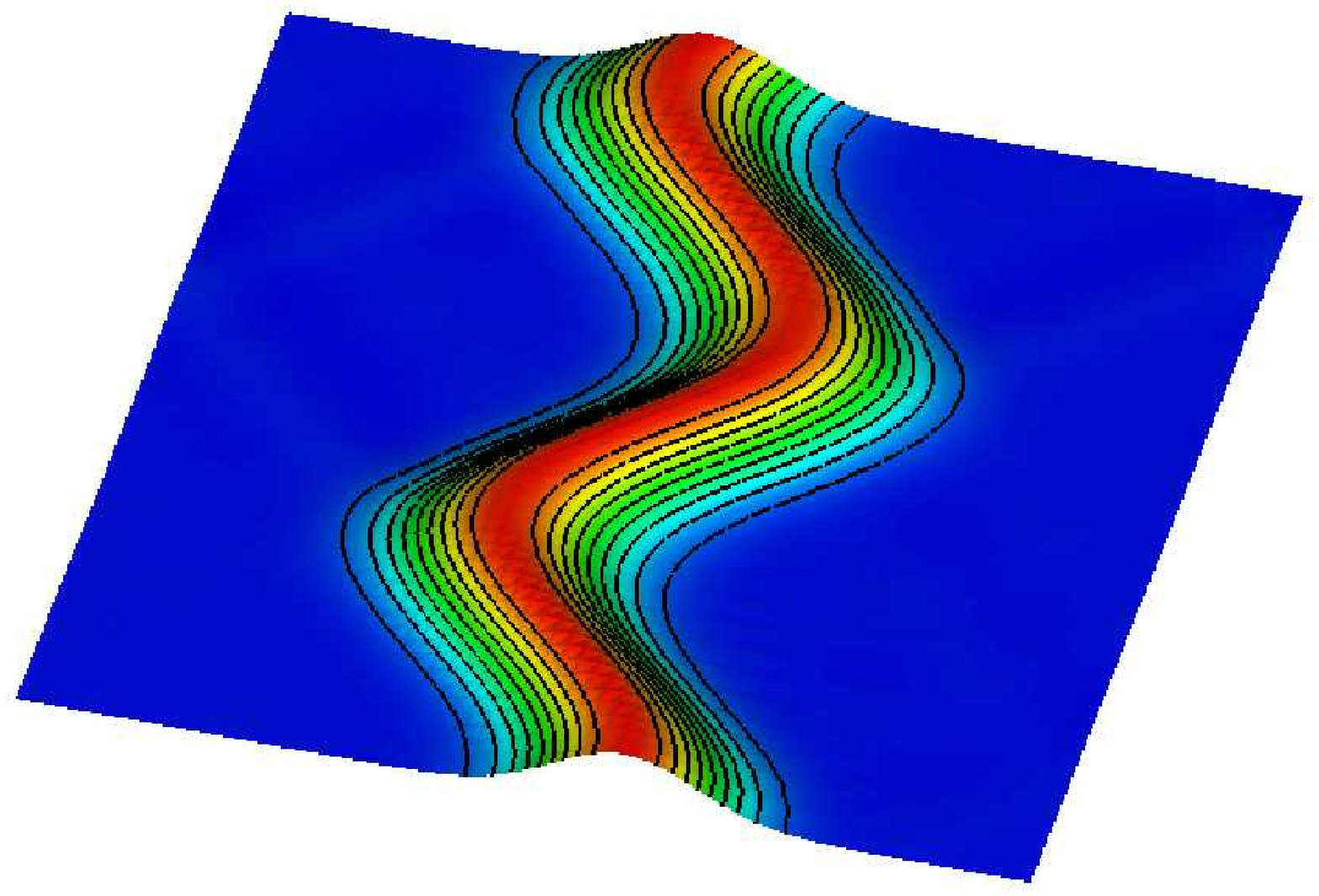,height=5.5cm}
\end{tabular}
\caption{An additive Schwarz preconditioned example using twelve subdomains in time on a $60^2$ structured mesh, referenced in Table \ref{table:1d_more}.
The plots show the solution after 1,10,100, and 500 GMRES iterations.  Unlike time marching
methods, the solution is constructed at all times simultaneously.  The preconditioner
substantially speeds up this process; evidence of the twelve additive Schwarz time subdomains is apparent after
the first iteration of GMRES.}
\label{fig:steps}
\end{figure}

Results for time decomposition of a $60^2$ mesh are summarized in Table \ref{table:1d_more}.
\begin{table}
\begin{tabular}{c|c|c|c|c|c}
Solver Type & Preconditioner & \# of subdomains & iterations & Final Residual & $\| \left(\tilde{u}-U_e\right) \|_{L_\infty}$ \\
  \hline
GMRES & ASM & 4 &  100 & $10^{-4}$ & $2.15 \times 10^{-2}$ \\
GMRES & ASM & 4 &  500 & $10^{-5}$ & $2.21 \times 10^{-2}$ \\
GMRES & ASM & 5 &  200 & $10^{-4}$ & $2.22 \times 10^{-2}$ \\
GMRES & ASM & 6 &  200 & $10^{-4}$ & $2.50 \times 10^{-2}$ \\
GMRES & ASM & 10 & 500 & $10^{-4}$ & $2.27 \times 10^{-2}$ \\
GMRES & ASM & 12 & 500 & $10^{-4}$ & $2.39 \times 10^{-2}$ \\
\end{tabular}
\caption{GMRES results using additive Schwarz method (ASM) preconditioning with a time decomposition strategy
  for the $1+1$ dimension case using a $60^2$ structured mesh.  
    The column labeled ``Final Residual" gives the absolute residual norm for the
      linear solve.
    All the results
  show significant improvement over the comparable unpreconditioned GMRES case shown in 
    Table \ref{table:1d}.  Increasing the number
    of time subdomains generally requires more GMRES iterations to achieve comparable
  error residuals; however, the preconditioner is potentially faster with more subdomains.
  Also, the preconditioner would be more scalable in parallel when using more subdomains.}
\label{table:1d_more}
\end{table}
Figure \ref{fig:steps} shows plots of the solution after 1, 10, 100, and 500 
GMRES iterations for the twelve subdomain additive Schwarz
case.  Subdomains were defined for this case by equally dividing up 
the global domain into time slabs
consisting of 5 or 6 nodes each in the time direction.

The additive Schwarz preconditioner gives excellent performance compared to GMRES alone and provides a scalable
alternative to LU decomposition for large problems.  Furthermore, the additive Schwarz preconditioner 
is already suitable for time-parallel computation; each processor would take a portion
of the time subdomain in preconditioning.  
Spatial domain decomposition could also be
explored in connection with time decomposition; however, we restrict our attention to time
decomposition here.

\subsection{$2+1$ Dimensions}
For $2+1$ dimensions, we modify the solution to be
\begin{eqnarray}
U_e \left(x,y,t\right) = \exp\left[-\left(x-\cos t\right)^2 - \left(y+\sin t\right)^2 \right].\label{eqn:2+1}
\end{eqnarray}
on a domain of $x,y = \left[-4,4\right]$ and $t=\left[0,4\right]$.  The linear system is constructed
using linear hexahedral elements giving second order convergence for the system.

A time decomposition strategy for preconditioning is also explored in $2+1$.  Like the $1+1$ case, 
  employing a time decomposition strategy with additive Schwarz preconditioning
significantly improves performance when compared to using GMRES alone or LU decomposition.
Table \ref{table:2d} gives a summary of results obtained using a $40 \times 40 \times 20$ mesh.
Performance times given are the solve times obtained on AMD opteron 250 processor with a clock
speed of 2.4 GHz using the PETSc timing utility.
\begin{table}
\begin{tabular}{c|c|c|c|c|c|c}
Solver Type & Preconditioner & \# of subdomains & iterations & Final Residual & $\| \left(\tilde{u}-U_e\right) \|_{L_\infty}$ & Time (sec) \\
  \hline
LU    &  -   & - &  -   & $10^{-14}$ & $2.73 \times 10^{-2}$ & $2.1 \times 10^{3}$\\
GMRES & none & 1 & 3000 & $10^{-2}$  & $2.10 \times 10^{-1}$ & $2.8 \times 10^{3}$\\
GMRES & ASM  & 4 & 500  & $10^{-4}$  & $2.92 \times 10^{-2}$ & $1.5 \times 10^{2}$\\
GMRES & ASM  & 4 & 1000 & $10^{-5}$  & $2.77 \times 10^{-2}$ & $4.0 \times 10^{2}$\\
GMRES & ASM  & 5 & 500  & $10^{-4}$  & $2.90 \times 10^{-2}$ & $1.5 \times 10^{2}$\\
GMRES & ASM  & 5 & 1000 & $10^{-4}$  & $2.77 \times 10^{-2}$ & $4.1 \times 10^{2}$\\
\end{tabular}
\caption{Linear solve results
  for the $2+1$ dimension case using a $40 \times 40 \times 20$ structured mesh.  
    The column labeled ``Final Residual" gives the absolute residual norm for the
      linear solve.
    The additive
  Schwarz preconditioned cases 
  show significant performance gain over the comparable unpreconditioned GMRES case and LU case.
  The performance times, given in seconds, were obtained using the PETSc timing utility running on an AMD
opteron 250 processor.  The GMRES simulations required significantly less memory than LU decomposition.
Figure \ref{fig:2p1_iter} shows the solution for the five subdomain ASM case after 10, 50, 100, and 500 GMRES iterations.}
\label{table:2d}
\end{table}

\begin{figure}
\begin{tabular}{cc}
After 10 iterations & After 50 iterations \\
\epsfig{file=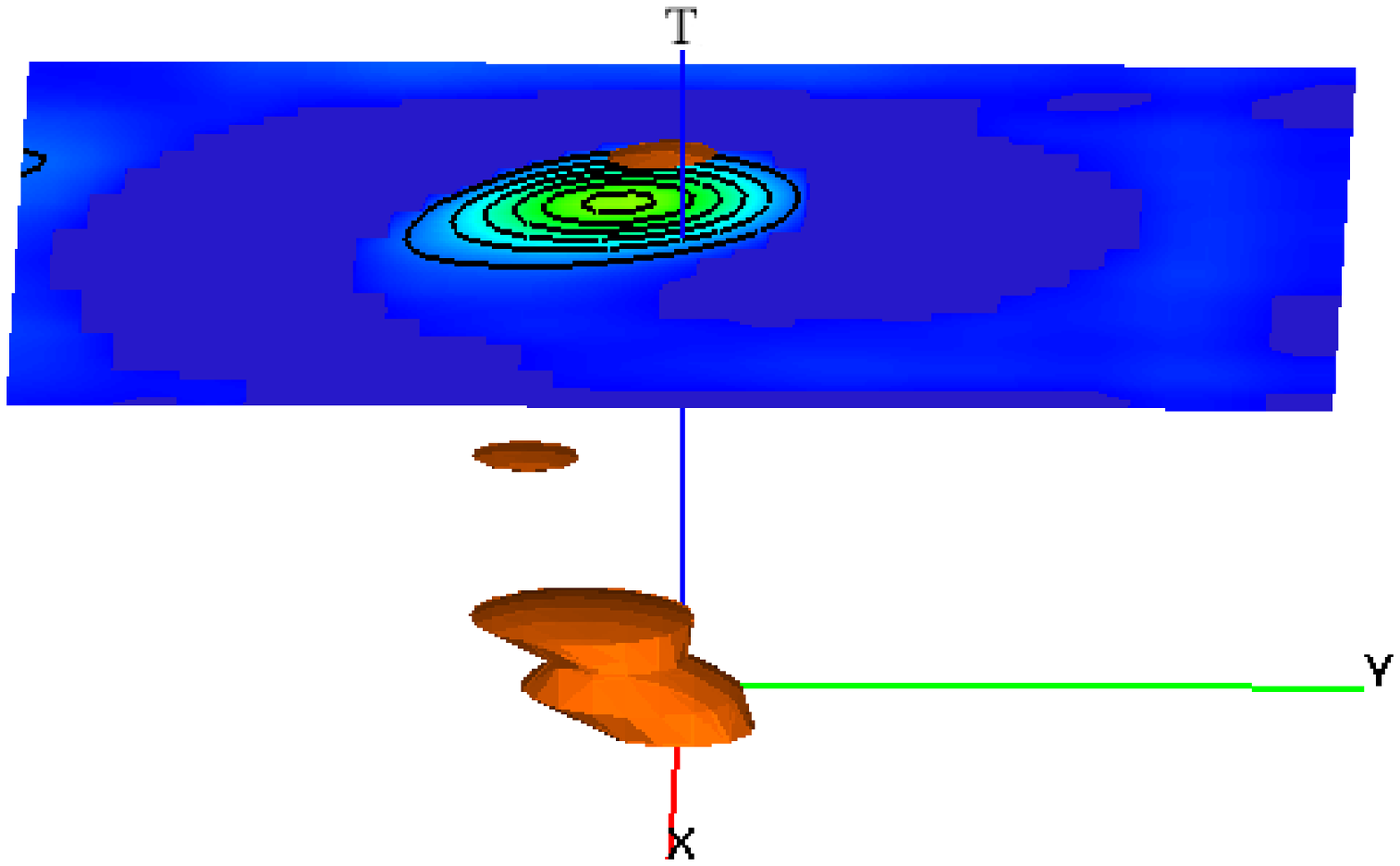,height=4.0cm} & \epsfig{file=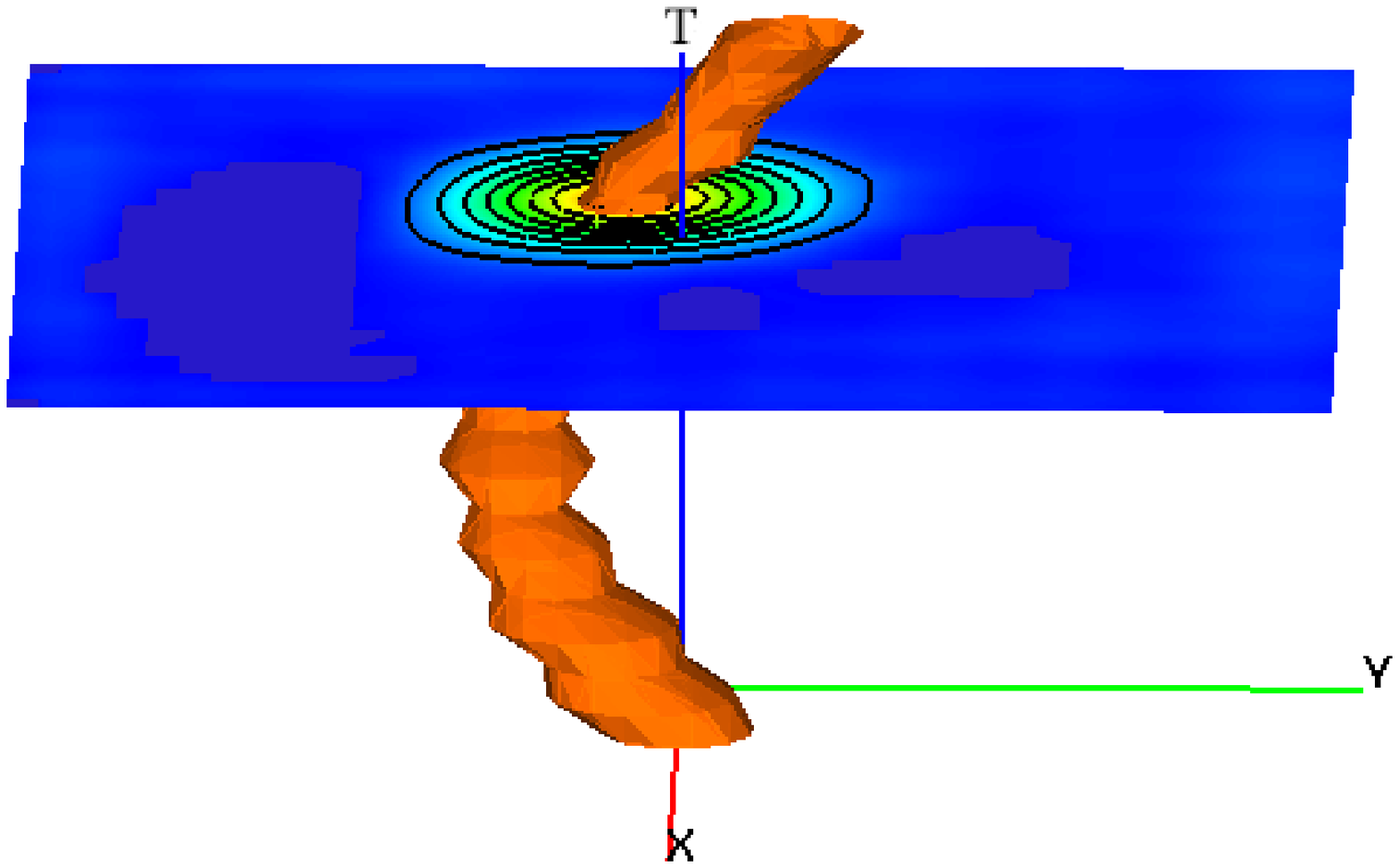,height=4.0cm} \\
 &  \\
After 100 iterations & After 500 iterations \\
\epsfig{file=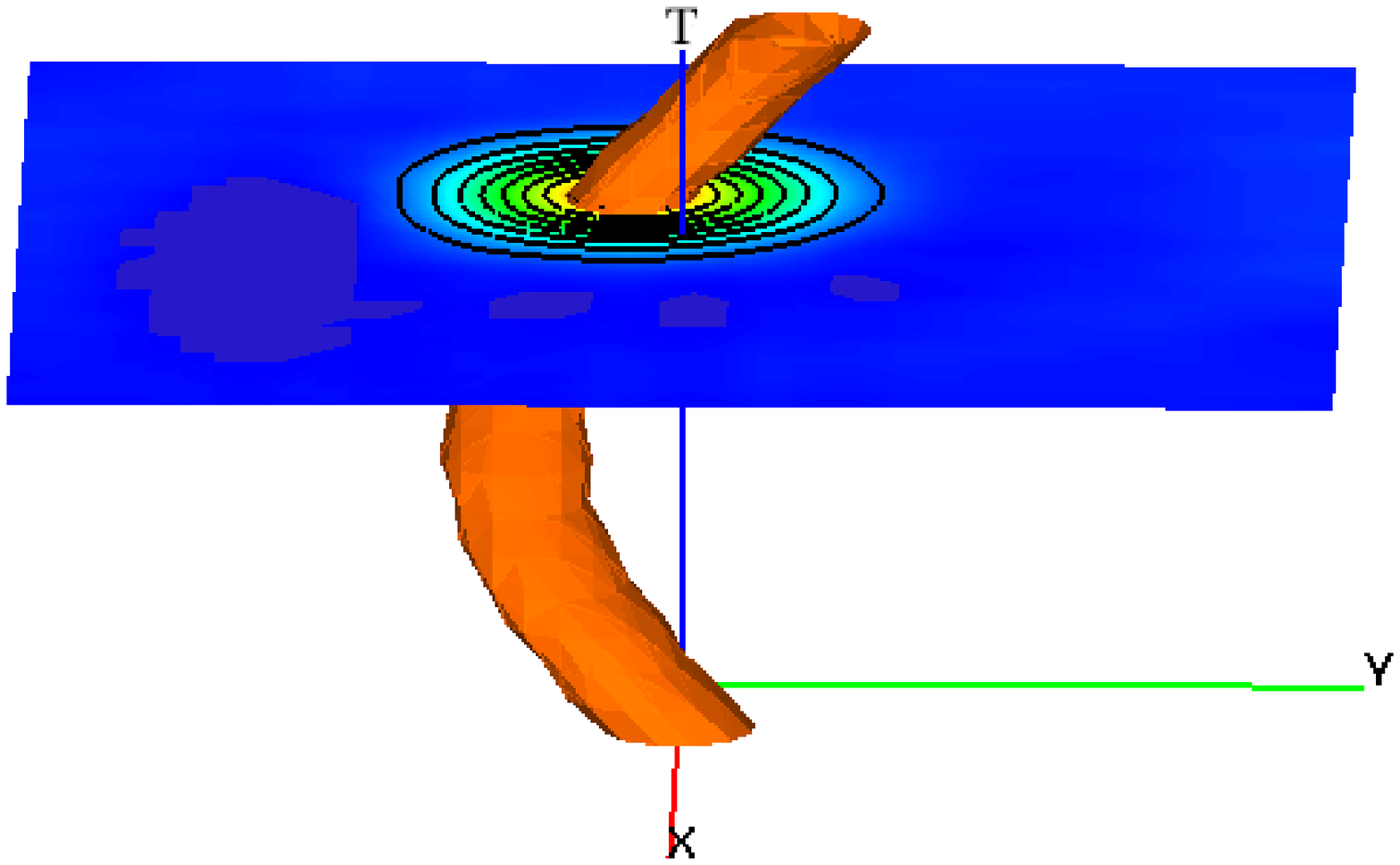,height=4.0cm} & \epsfig{file=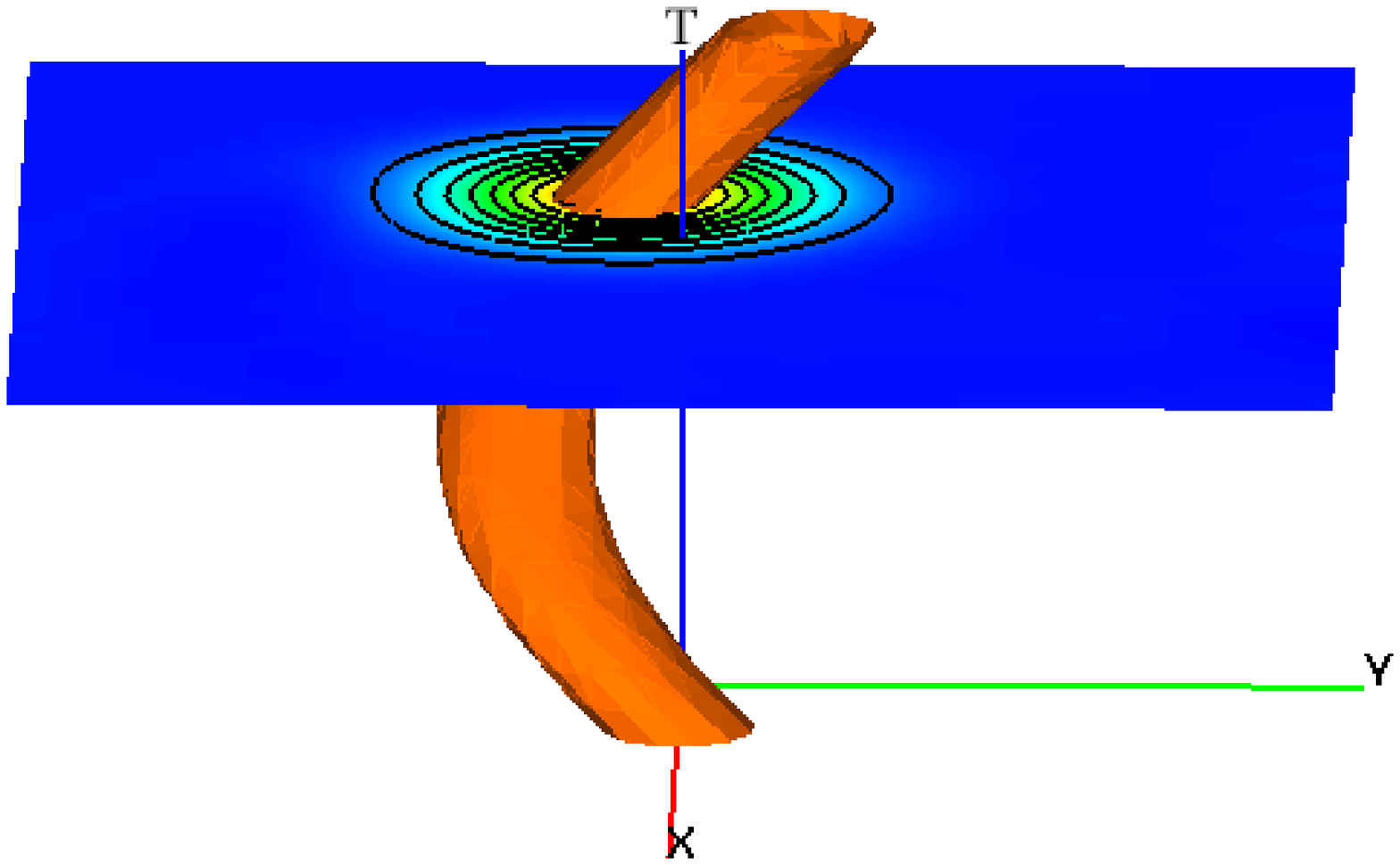,height=4.0cm} 
\end{tabular}
\caption{An additive Schwarz preconditioned example in $2+1$ dimensions using five subdomains in time on a $40 \times 40 \times 20$ structured mesh, referenced in Table \ref{table:2d}.
The plots show the solution after 10,50,100, and 500 GMRES iterations.  
  The isosurface indicates a surface with value of 0.8, tracking the motion of the pulse in time.  A slice
 of the solution at time 3 is also shown with contour lines on the slice.  The vertical axis is the time direction.  
 Like Figure \ref{fig:steps} in the $1+1$ dimension case,
the solution is constructed at all timesteps at once rather than sequentially solving a single timestep at a time
  as in time marching methods}.  
\label{fig:2p1_iter}
\end{figure}

GMRES without preconditioning is ineffective for this problem due to the slow convergence rate.
As expected, LU decomposition is also ineffective due to poor scaling as the problem size grows.  
In contrast, GMRES with additive Schwarz method (ASM) preconditioning
using a time decomposition strategy is significantly more effective.
Figure \ref{fig:2p1_iter} shows plots of the solution after 10, 50, 100, and 500 GMRES iterations for the five subdomain
ASM preconditioned case.  As in the $1+1$ cases, the ASM preconditioner is time-parallel: parallelization can be 
achieved by simultaneously solving each time subdomain on a different processor.

\subsection{$3+1$ Dimensions}
\begin{figure}
\epsfig{file=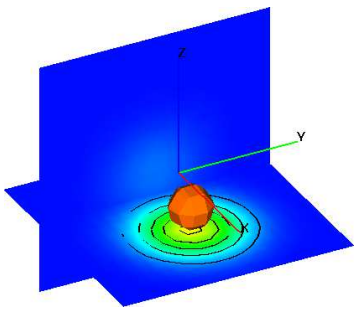,height=4.0cm}
\caption{Initial data for $3+1$ simulation in Figure \ref{fig:3p1}.}
\label{fig:3p1_initial}
\end{figure}

\begin{figure}
\begin{tabular}{cccc}
               & Time 1: & Time 3: & Time 5: \\
10 iterations: & \epsfig{file=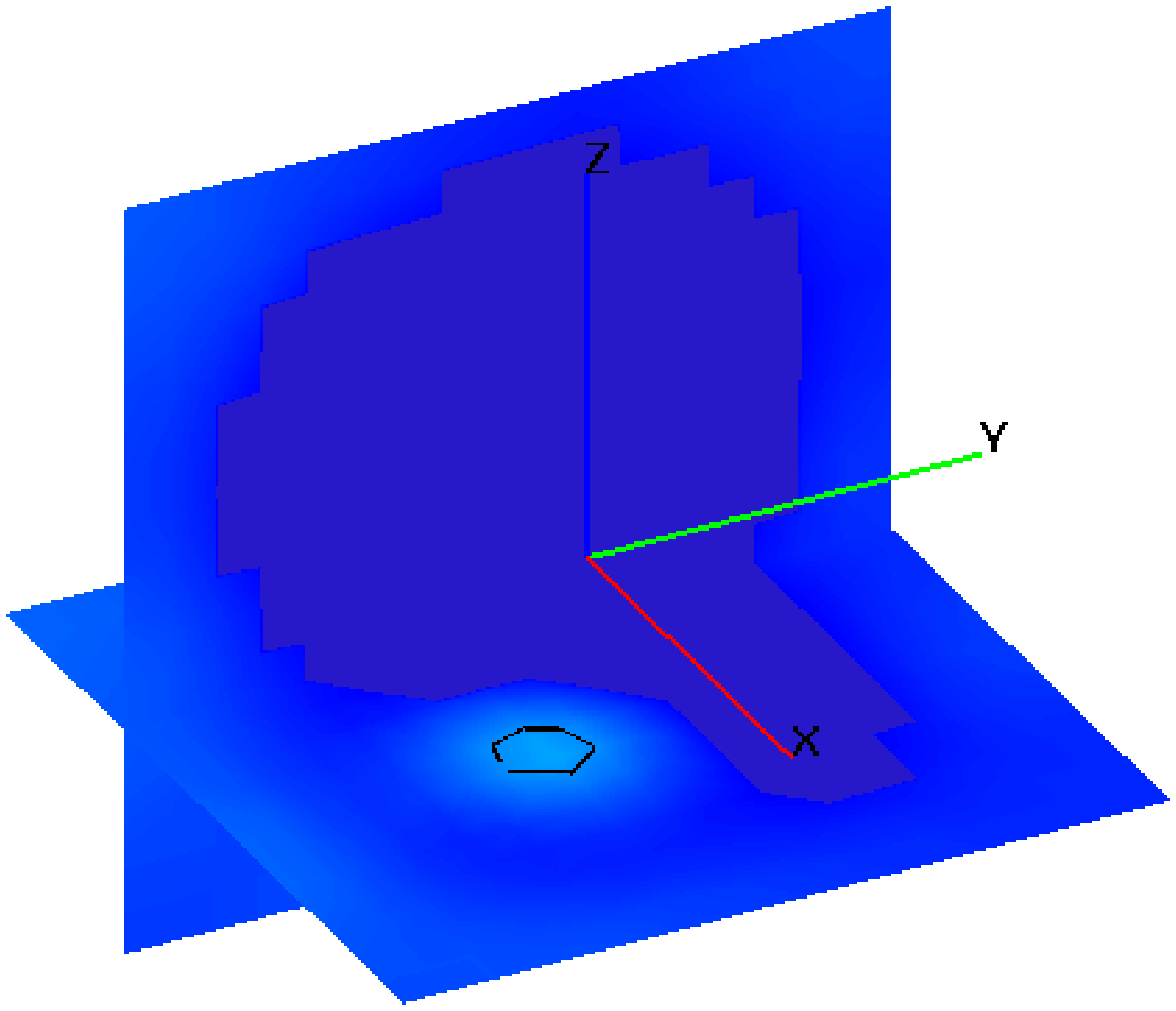,height=4.0cm} & \epsfig{file=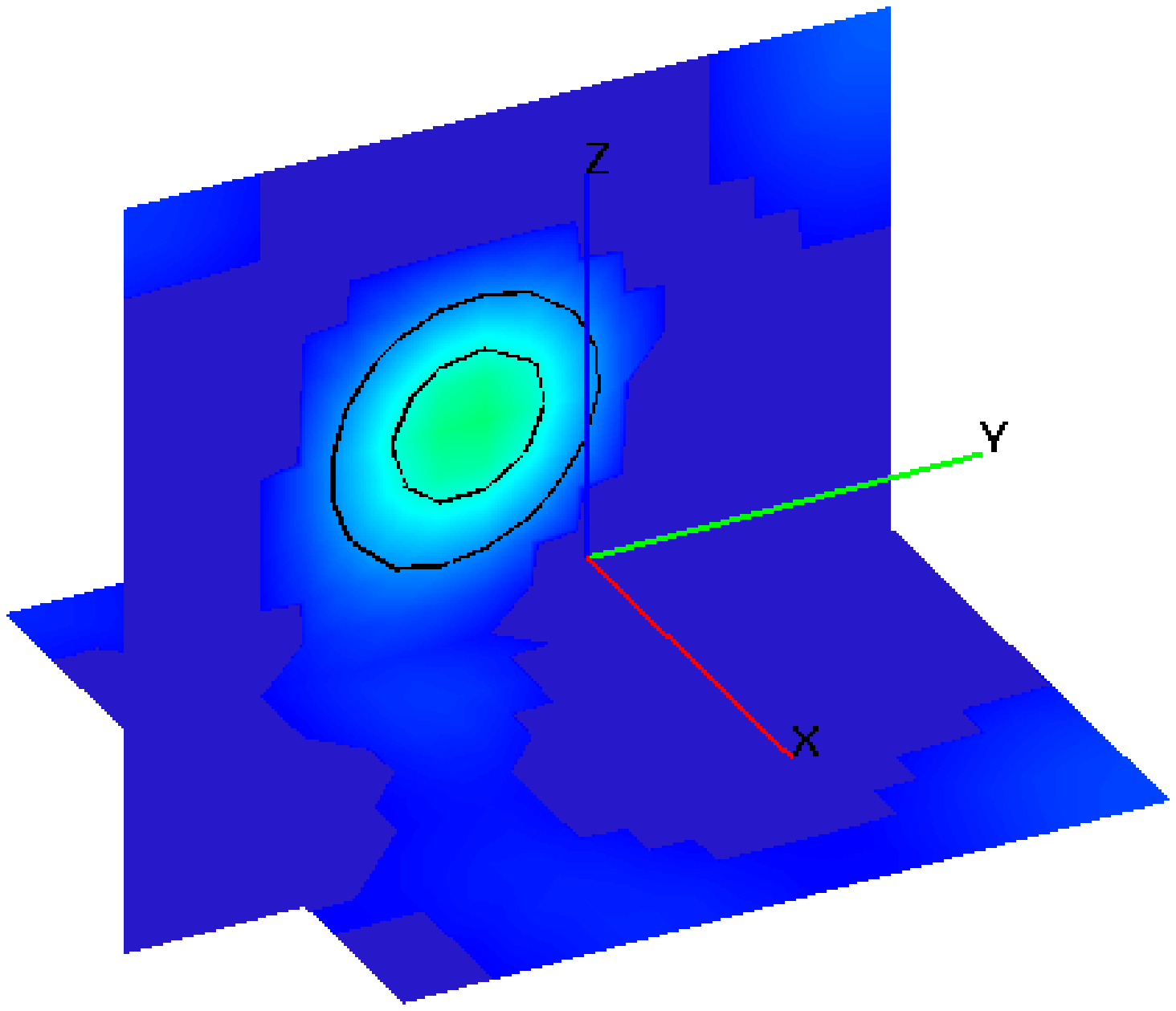,height=4.0cm} &
\epsfig{file=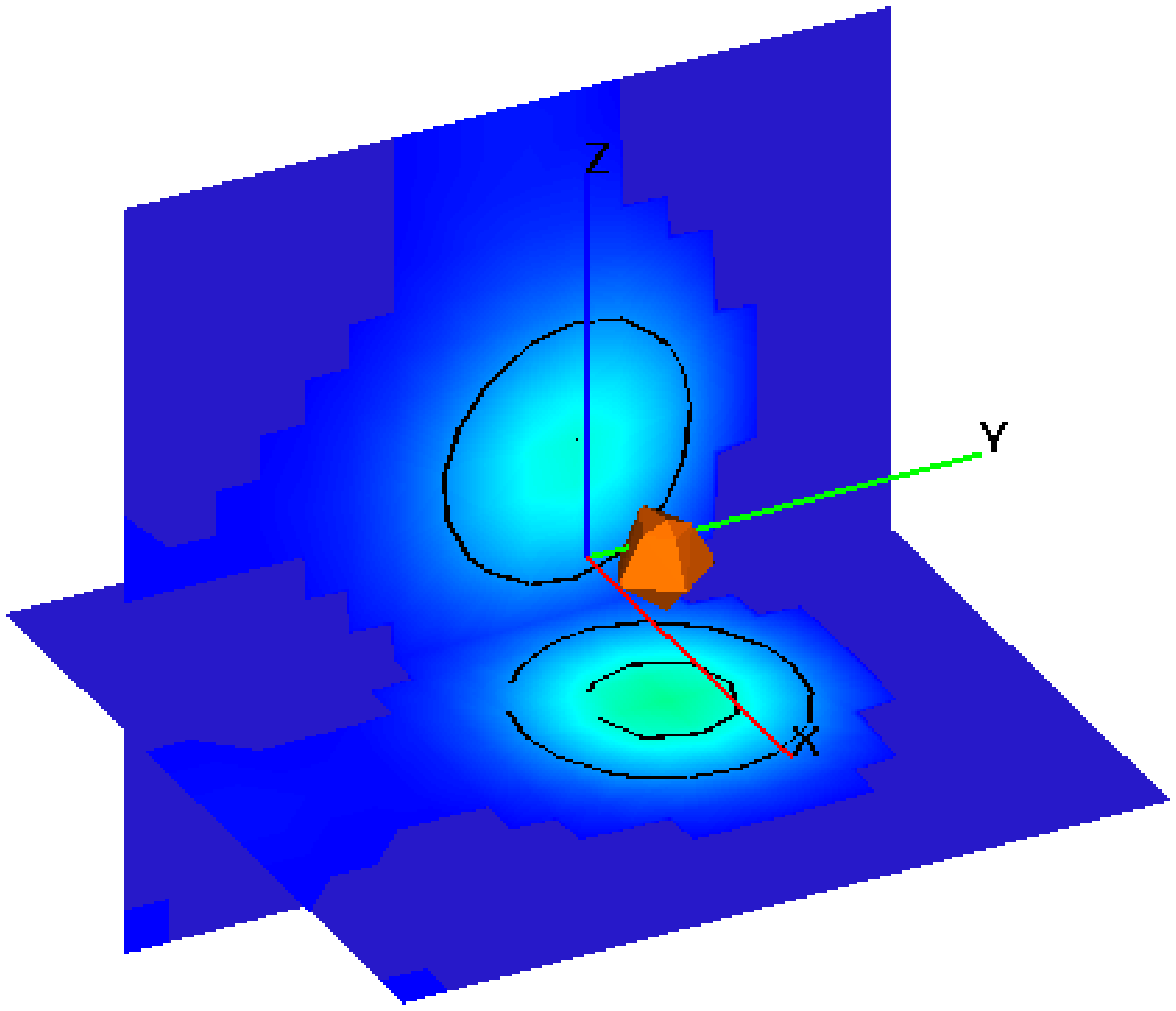,height=4.0cm} \\
50 iterations: & \epsfig{file=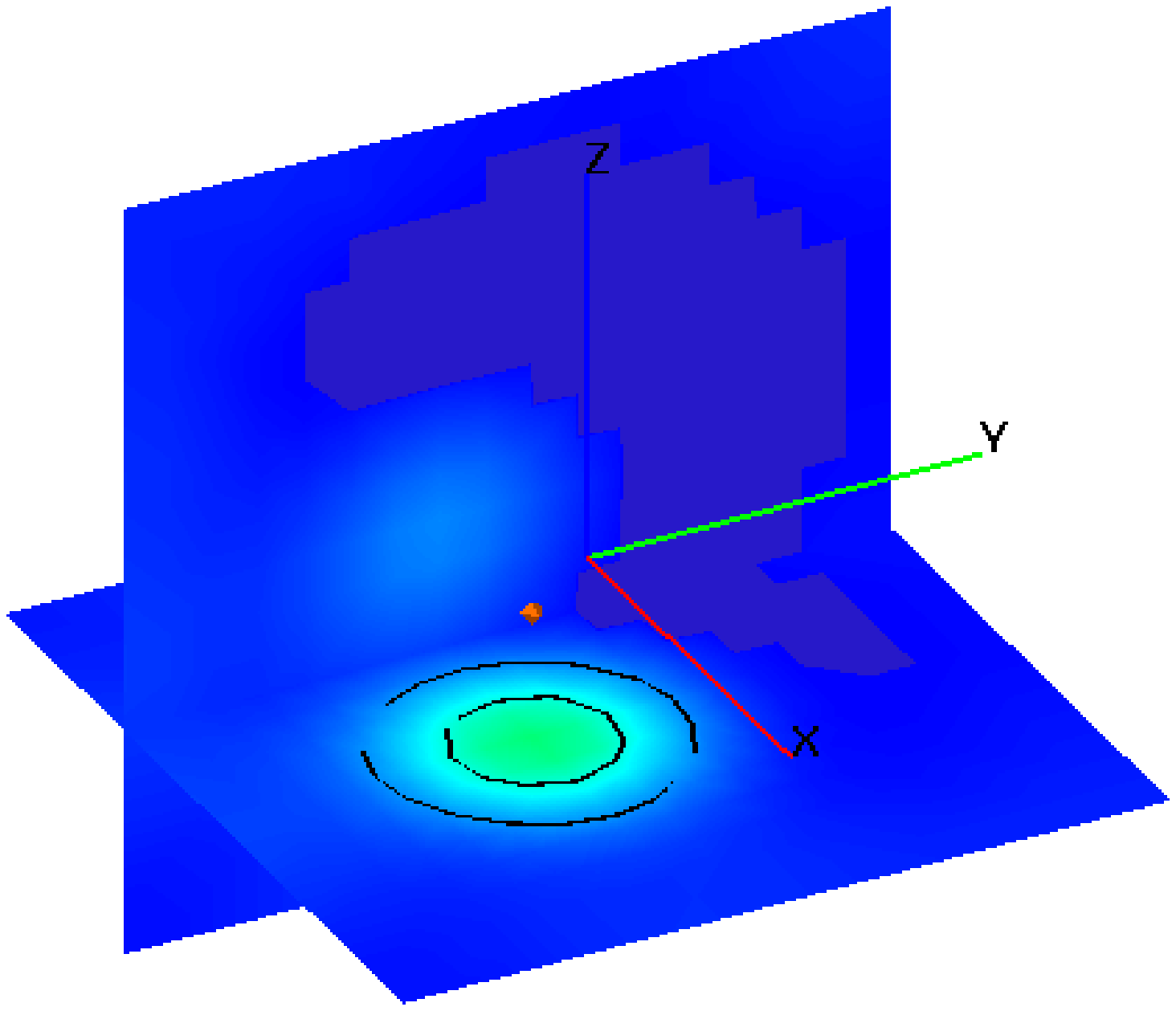,height=4.0cm} & \epsfig{file=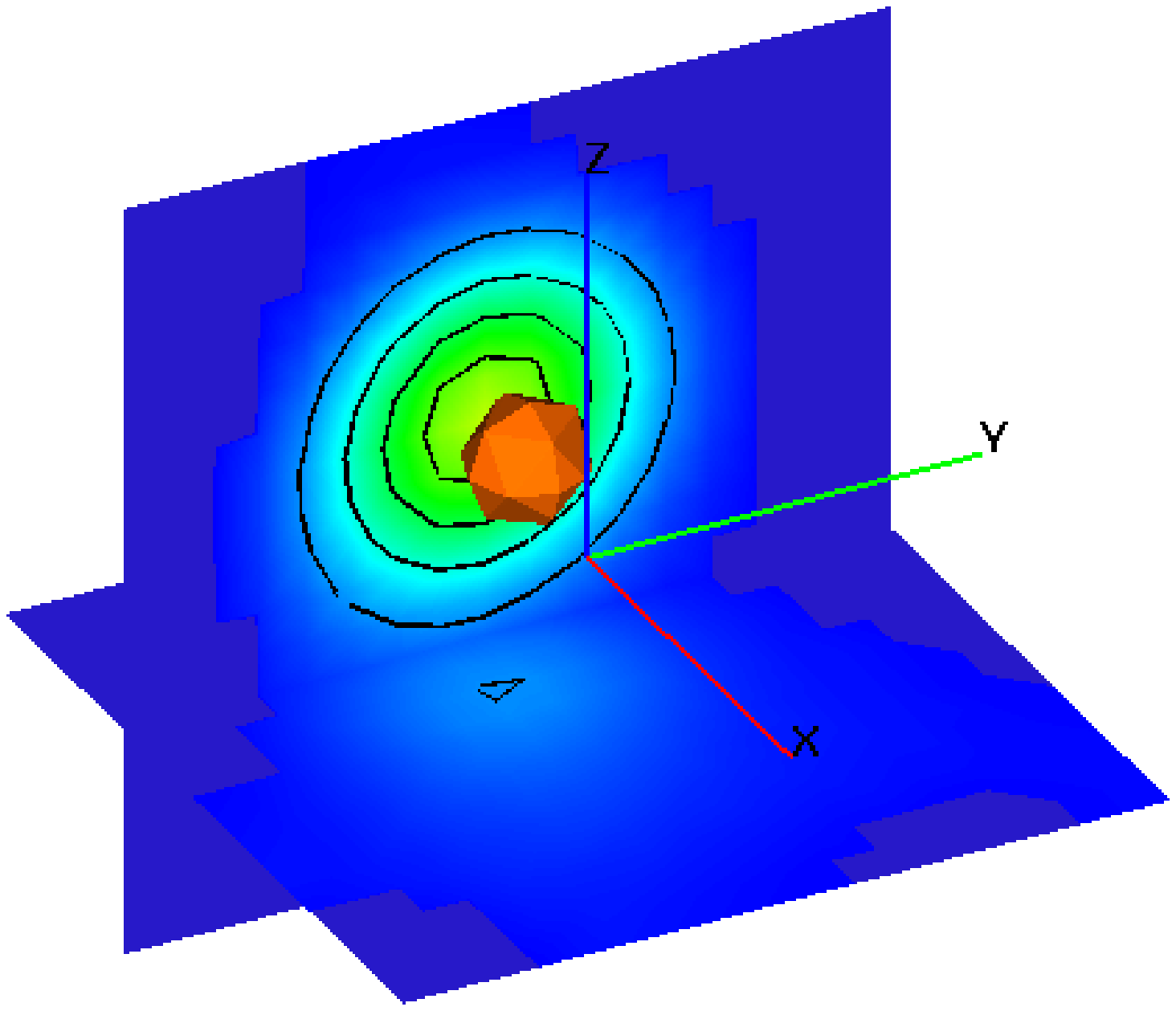,height=4.0cm} &
\epsfig{file=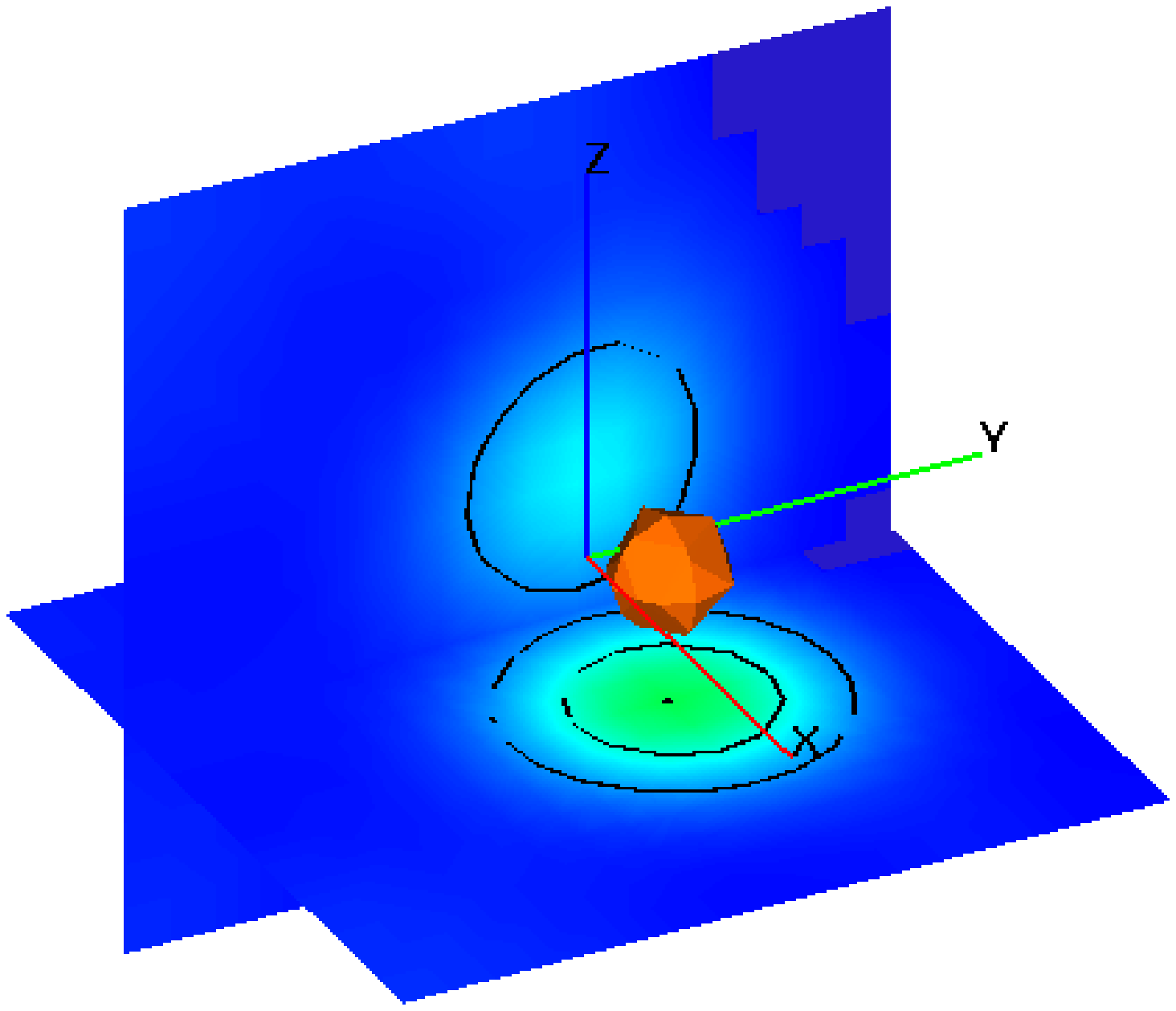,height=4.0cm} \\
100 iterations: & \epsfig{file=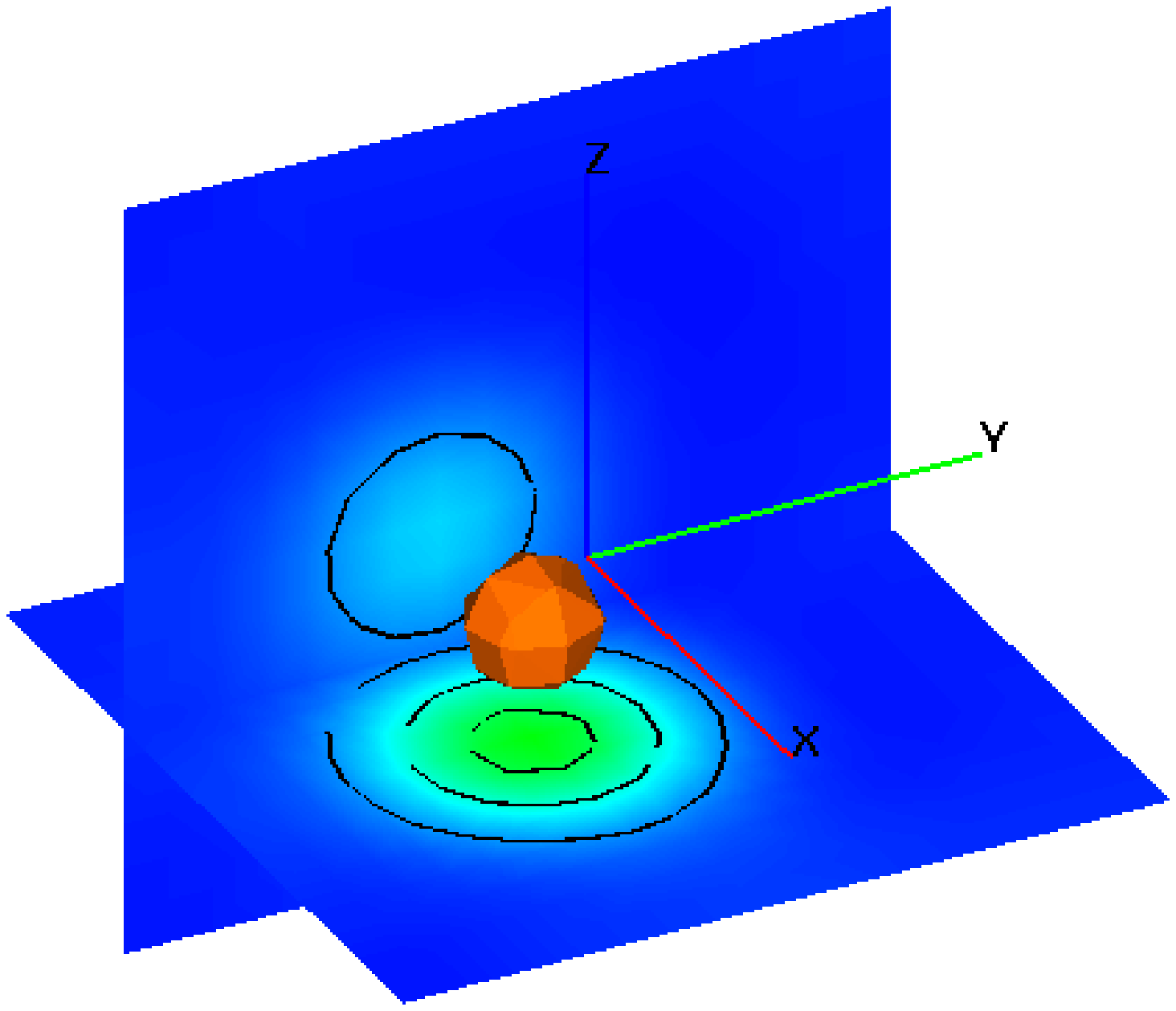,height=4.0cm} & \epsfig{file=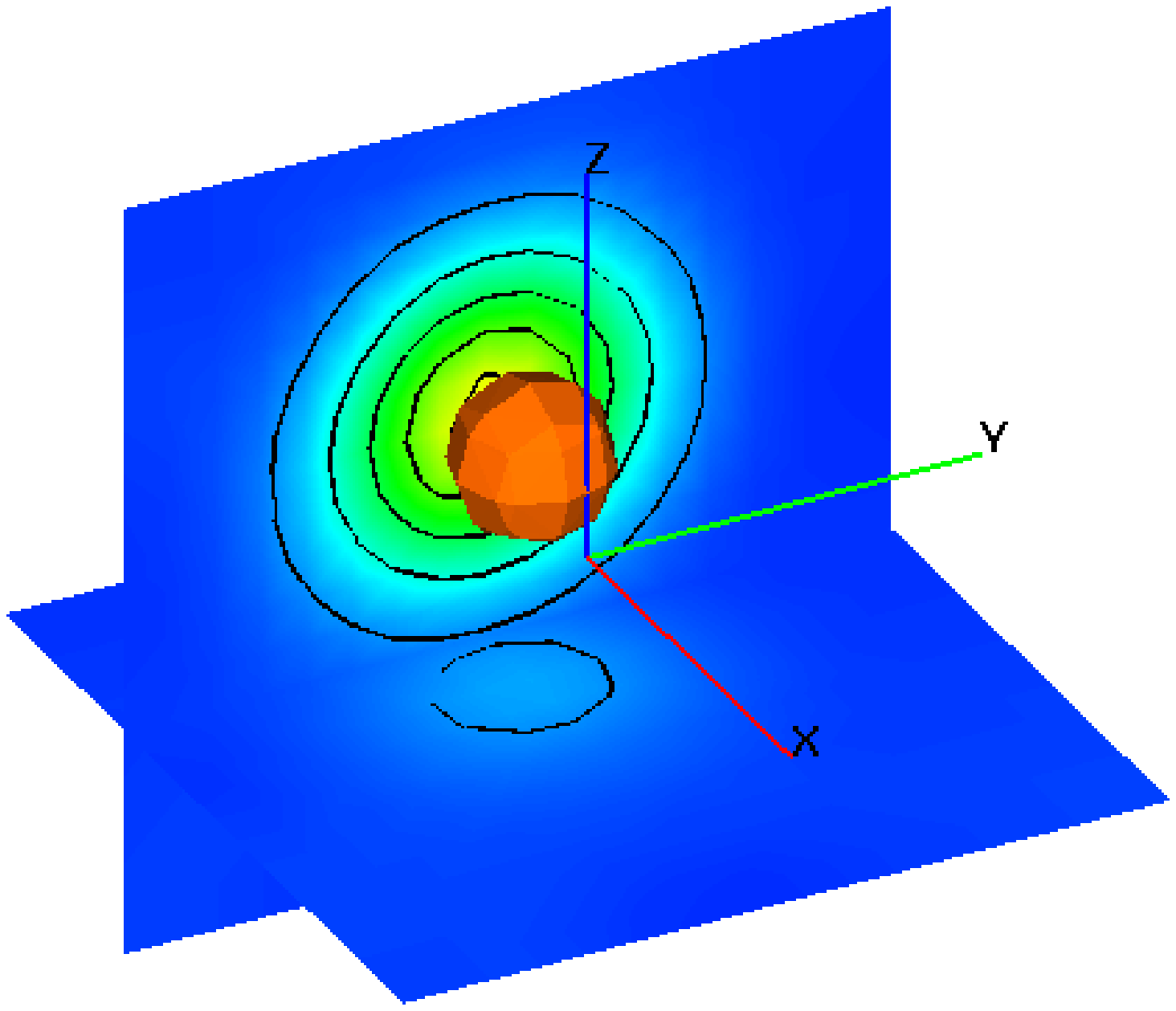,height=4.0cm} &
\epsfig{file=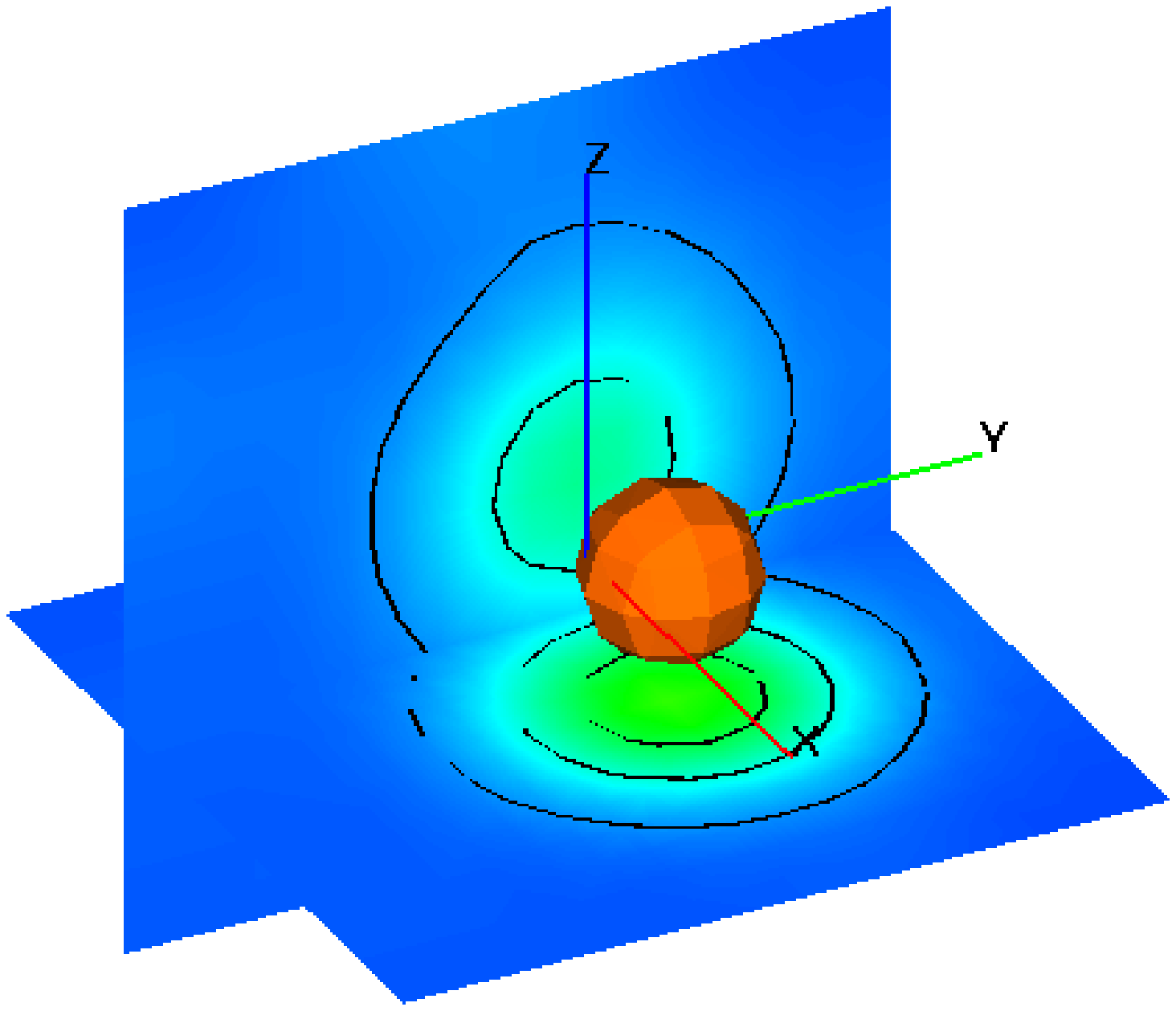,height=4.0cm} \\
750 iterations: & \epsfig{file=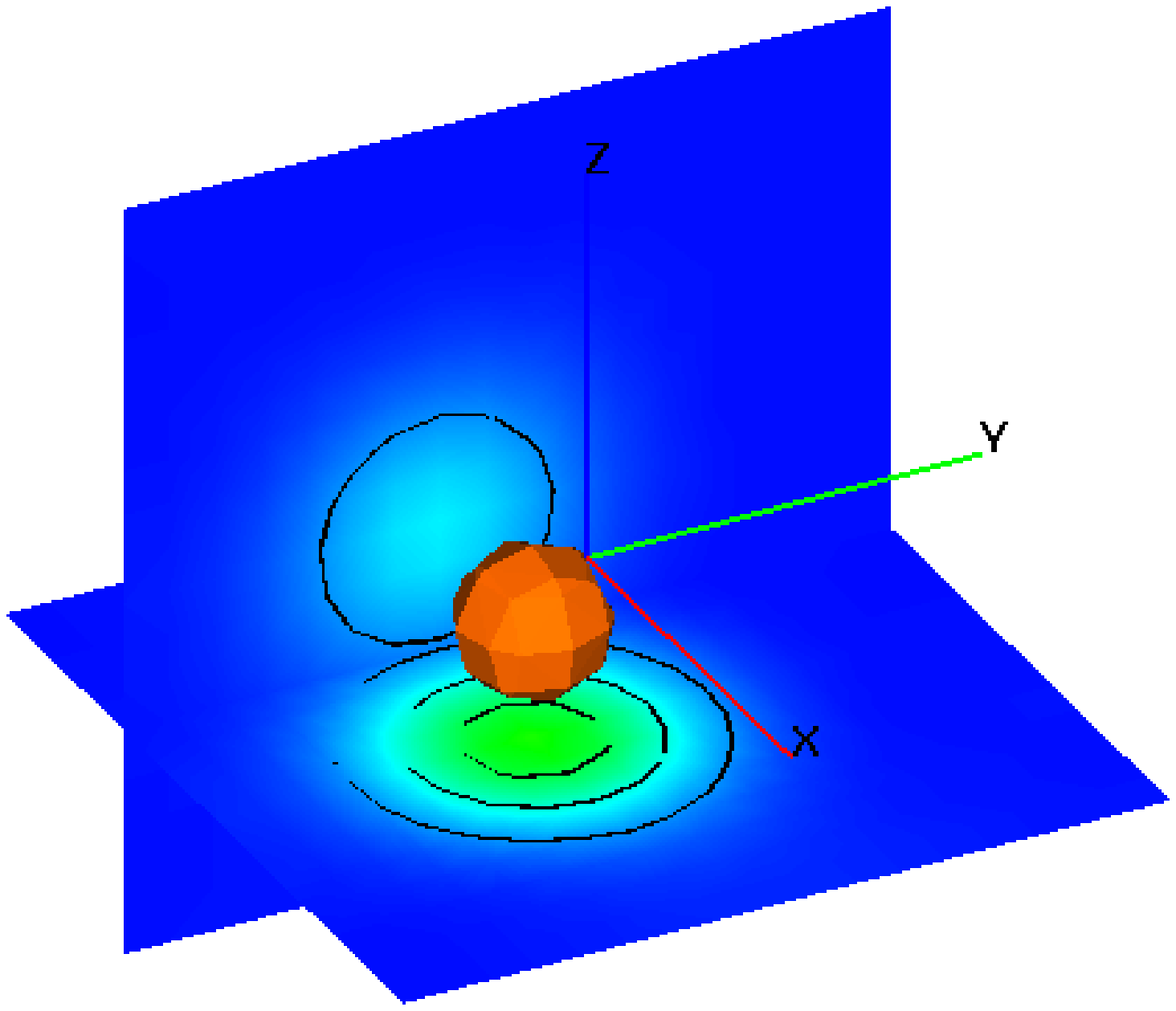,height=4.0cm} & \epsfig{file=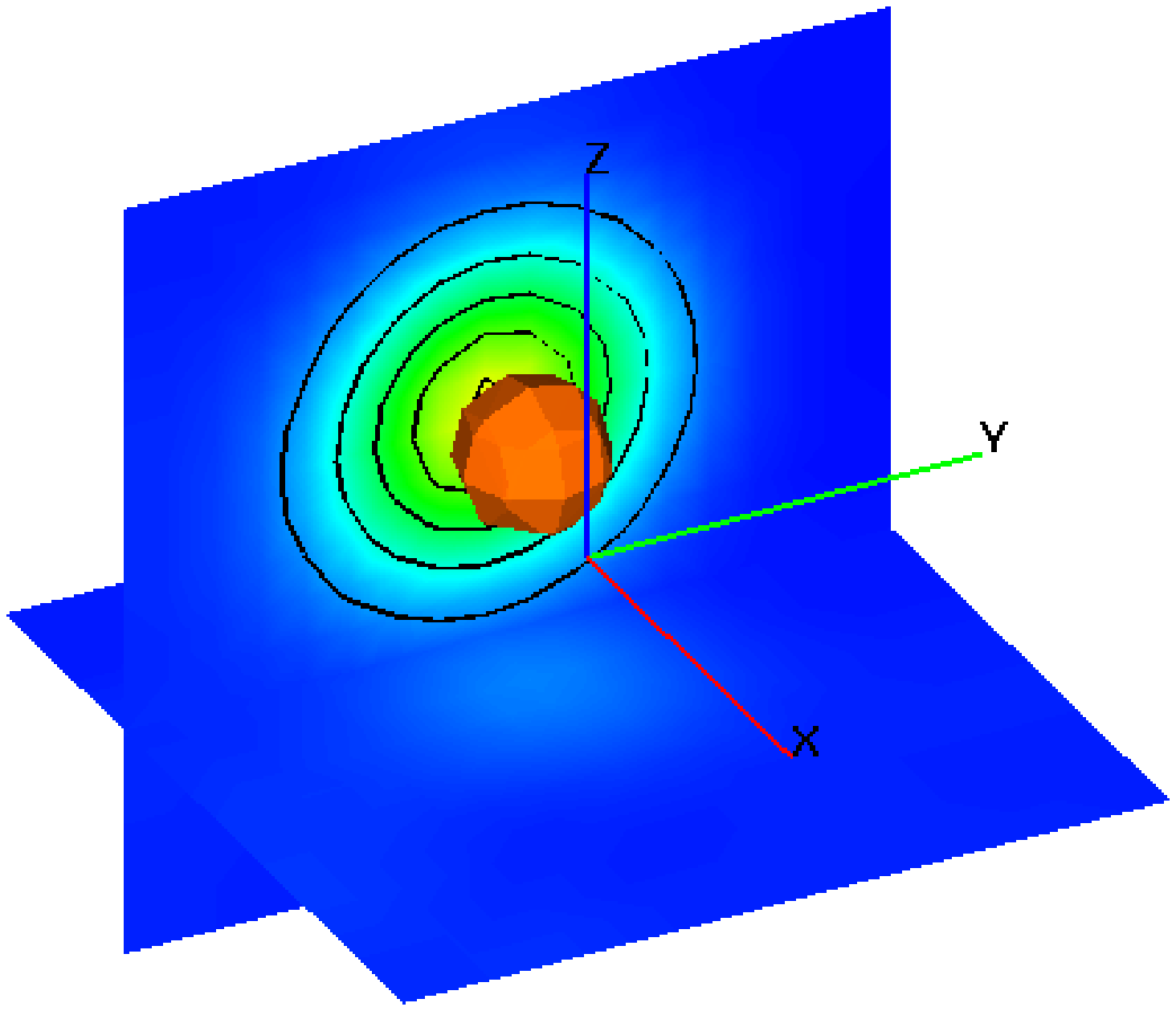,height=4.0cm} &
\epsfig{file=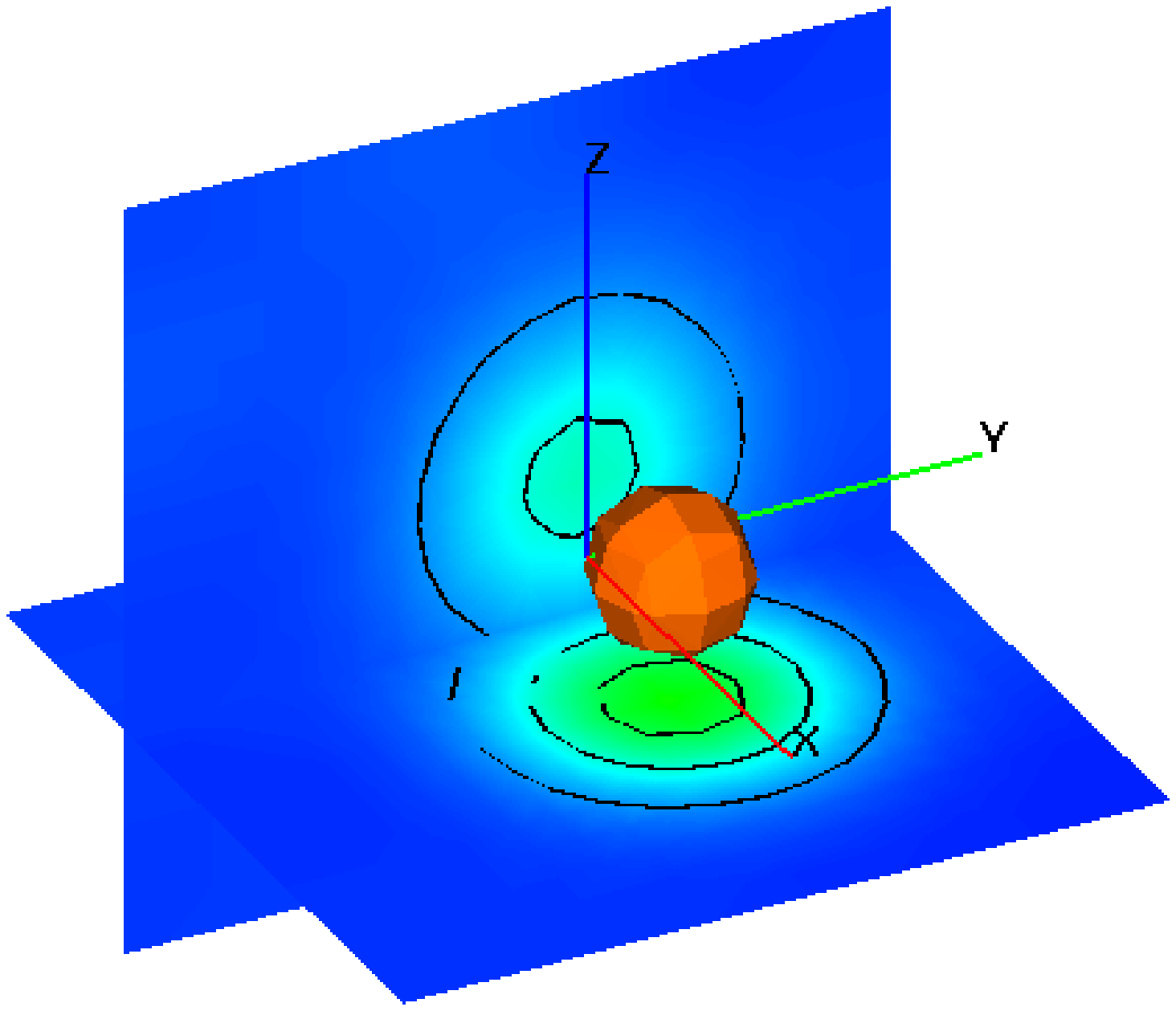,height=4.0cm}
\end{tabular}

\caption{The additive Schwarz preconditioned $3+1$ example using nine subdomains in time on a $18^4$ structured mesh
  referred to in Table \ref{table:3d}.  The initial data for this simulation is shown in Figure \ref{fig:3p1_initial}.
The plots here show the solution, by row, after 10,50,100, and 750 GMRES iterations.  The columns indicate
the time of the solution:
  the left column shows 
the solution at time 1, the middle column shows time 3, and the right column shows time 5.
An isosurface with value 0.8 tracks the motion of the pulse.  Two spatial planes with contours are
also shown: the first on the $x=-1$ plane, and the second on the $z=-1$ plane.  
Like Figures \ref{fig:steps} and \ref{fig:2p1_iter}, the solution 
is constructed at all times simultaneously.  The mesh consists of tesseracts, the
higher dimensional analogue of hexahedra.  The 3-D datasets shown are time slices from the 4-D mesh.  
}
\label{fig:3p1}
\end{figure}
For $3+1$ dimensions, we select the solution to be
\begin{eqnarray}
U_e \left(x,y,z,t\right) = \exp\left[-\left(x-\frac{1}{2} \cos t\right)^2 - \left(y+\frac{1}{2}\sin t\right)^2 - \left(z+\frac{1}{2}\cos t\right)^2 \right].\label{eqn:3+1}
\end{eqnarray}
on a domain of $x,y,z = \left[-2.5,2.5\right]$ and $t=\left[0,5\right]$.  The linear system is constructed
using linear tesseractic elements consisting of 16 nodes per element,
giving second order convergence for the system.
Tesseracts are the higher dimensional analogue of hexahedra \cite{Mathworld}.  
Table \ref{table:3d} gives a summary of results obtained using a $18^4$ mesh.

\begin{table}
\begin{tabular}{c|c|c|c|c|c|c}
Solver Type & Preconditioner & \# of subdomains & iterations & Final Residual & $\| \left(\tilde{u}-U_e\right) \|_{L_\infty}$ \\
  \hline
GMRES & ASM  & 6 & 300  & $10^{-4}$  & $8.13 \times 10^{-2}$ \\
GMRES & ASM  & 9 & 300  & $10^{-3}$  & $8.53 \times 10^{-2}$ \\
GMRES & ASM  & 9 & 600  & $10^{-3}$  & $8.17 \times 10^{-2}$ \\
GMRES & ASM  & 9 & 750  & $10^{-4}$  & $8.15 \times 10^{-2}$ \\
\end{tabular}
\caption{Linear solve results
  for the $3+1$ dimension case using an $18^4$ structured mesh.  
    The column labeled ``Final Residual" gives the absolute residual norm for the
      linear solve.
  Unlike the $1+1$ and $2+1$ cases, these $3+1$ examples were only possible on a single processor because of the
  additive Schwarz preconditioning; both LU decomposition and unpreconditioned GMRES
  were impractical because of memory limitations or time limitations.
  These $3+1$ simulations used a mesh composed of linear tesseracts: 16 node 4-D hyperelements.
Figure \ref{fig:3p1_initial} shows the initial data used for these simulations; Figure \ref{fig:3p1} shows 
the solution for the nine subdomain ASM case after 10, 50, 100, and 750 GMRES iterations
at three times: 1,3, and 5.}
\label{table:3d}
\end{table}
Figures \ref{fig:3p1_initial}--\ref{fig:3p1} show plots of the solution
at selected time slices; Figure \ref{fig:3p1} shows the nine subdomain
additive Schwarz preconditioned case at 10, 50, 100, and 750 GMRES iterations.

\section{Conclusions}
We have numerically examined space-time finite elements for the nonhomogeneous wave equation,
  testing several types of linear solvers and preconditioners in several dimensions.  
  The motivation of this study is to explore the performance issues surrounding the
  use of space-time elements in the context of numerical relativity.
  Fully unstructured meshes in space and time can greatly simplify issues surrounding 
  time-varying computational domains and space-time mesh refinement, 
  provided that both the domain and refinement are specified a priori.  They have also shown
  promise when the time-varying domain is not known a priori, as in \cite{Walhorn} and \cite{Hansbo}.
  We restricted our attention to those simulations which could be performed on a single processor.
Fully implicit examples using a continuous time Galerkin method were presented
in $1+1$, $2+1$, and $3+1$ dimensions using linear quadrilateral, hexahedral, and tesseractic elements.

We found that LU decomposition and unpreconditioned GMRES were both capable of solving the linear systems which appear
in these space-time element simulations.  However, both choices scaled too poorly with respect to problem size to be
effective even for moderate size simulations in $3+1$.
Standard preconditioners like Jacobi and Block-Jacobi did not improve GMRES performance for the space-time
linear systems.

We found that additive Schwarz preconditioning significantly improved GMRES performance.  Substantial performance 
improvements were observed by applying a time decomposition strategy in additive Schwarz preconditioning.
The time decomposition strategy consisted of decomposing the global mesh into several smaller time subdomains for use
in preconditioning.  This preconditioning strategy is also time-parallel: all the time subdomains used in
preconditioning can be solved simultaneously on separate processors.  

Several improvements upon the additive Schwarz preconditioner remain to be explored.  In the experiments presented
here, only face cell overlap was examined.  Also, no attempt was made to combine time decomposition with spatial
domain decomposition even though such a combination would be natural. A study of the optimal interface condition 
\cite{kimn:2005} is another interesting question since
the interface condition explored here was physically motivated.
Attempts at a parallel implementation of the preconditioner will be forthcoming.  The substantial
performance benefits of the ASM preconditioner make further study into space-time elements for numerical
relativity feasible.

\section{Acknowledgements}
Early prototyping benefited from two finite element packages: 
libmesh \cite{libmesh} and Diffpack \cite{diffpack}.  
We also acknowledge helpful discussions with Luis Lehner and Jorge Pullin.
This work was supported by the following grants:
NSF-PHY-0244335, NSF-PHY-0244299, NSF-INT0204937, and NASA-NAG5-13430.

\bibliography{./stasm}
\bibliographystyle{unsrt}

\end{document}